\documentclass[useAMS,usenatbib]{mn2e}

\usepackage{pifont}
\usepackage[final]{graphicx}
\usepackage{amssymb}
\usepackage{color}
\usepackage{url}

\title[IR emission from central obscuring
structures]{Time-resolved infrared emission from radiation-driven
 central obscuring structures in Active Galactic Nuclei}
\author[M. Schartmann et al.]
  {M.~Schartmann$^{1,2,}$\thanks{E-mail: schartmann@mpe.mpg.de},
   K.~Wada$^{3}$,
   M.~A.~Prieto$^{4,5}$,
   A.~Burkert$^{1,2,}$\thanks{Max Planck Fellow}
   \newauthor
   and K.~R.~W.~Tristram$^{6}$\\
$^{1}$Universit\"ats-Sternwarte M\"unchen, Scheinerstra\ss e 1, D-81679 M\"unchen, Germany\\
$^{2}$Max-Planck-Institut f\"ur extraterrestrische Physik, Postfach 1312, Giessenbachstra\ss e, D-85741 Garching, Germany\\
$^{3}$Graduate School of Science and Engineering, Kagoshima
University, Kagoshima 890-0065, Japan\\
$^{4}$Instituto de Astrof\'{i}sica de Canarias (IAC), E-38200 La
Laguna, Tenerife, Spain\\
$^{5}$Universidad de La Laguna, Dept.~Astrof\'{i}sica, E-38206 La
Laguna, Tenerife, Spain\\
$^{6}$ESO -- European Organisation for Astron. Research in the Southern Hemisphere, Casilla 19001, Santiago, Chile
}

\begin{document}

\date{Accepted . Received ; in original form }

\pagerange{\pageref{firstpage}--\pageref{lastpage}} \pubyear{2014}

\maketitle

\label{firstpage}

\begin{abstract}
The central engines of Seyfert galaxies are thought to be enshrouded
by geometrically thick gas and dust structures. In this article, we
derive observable properties for a self-consistent model of such
toroidal gas and dust distributions,
where the geometrical thickness is achieved and maintained with the help of
X-ray heating and radiation pressure due to the central engine.
Spectral energy distributions (SEDs) and images are obtained with the
help of dust continuum radiative transfer calculations with RADMC-3D.
For the first time, we are able to present time-resolved SEDs and
images for a physical model of the central obscurer.
Temporal changes are mostly visible at shorter wavelengths, close to the
combined peak of the dust opacity as well as the central source
spectrum and are caused by variations in the column densities of the generated outflow. 
Due to the three-component morphology of the hydrodynamical models --
a thin disc with high density filaments, a
surrounding fluffy component (the obscurer) and a low density
outflow along the rotation axis -- 
we find dramatic differences depending on wavelength: whereas the
mid-infrared images are dominated by the elongated appearance of the
outflow cone, the long wavelength emission is mainly given by the cold
and dense disc component. Overall, we find good agreement with
observed characteristics, especially for those models, which show
clear outflow cones in combination with a geometrically thick
distribution of gas and dust, as well as a geometrically thin, but
high column density disc in the equatorial plane. 
\end{abstract}


\section{Introduction}
\label{sec:introduction}

Active Galactic Nuclei (AGNs) are thought to be powered by accretion onto a
supermassive black hole (SMBH, $10^6-10^{10}\,\mathrm{M}_{\odot}$). A large fraction of the gravitational energy
released by this process is emitted in form of radiation in the
ultraviolet (UV) and optical wavelength 
regime, originating
from a viscously heated accretion disc surrounding the SMBH \citep{Shakura_73}. A
fraction of the energy contained in this so-called {\it Big Blue Bump}
in the spectral energy distribution (SED) 
is absorbed and re-emitted in the infrared (IR) by dust \citep{Rees_69,Barvainis_87},
which partly enshrouds the central engine. The latter gives rise to another
characteristic feature in the SEDs of
AGN -- the so-called {\it IR bump}. 
The existence of overlaid broad emission lines (tracers of gas close
to the central engine) categorises an object
as a type~1 source, whereas galaxies with narrow emission lines only
are classified as type~2 sources. 
This observed dichotomy was suggested to be the result of a geometry
effect due to a toroidally shaped
absorber, usually referred to as the {\it dusty molecular torus}.
In type 1 sources the torus is thought to be seen face-on and in
type 2 sources edge-on, where it blocks the direct line of sight towards the
centre and only narrow emission lines from gas further out can be
detected. 
This is the essence of the so-called {\it Unified Scheme of Active
Galactic Nuclei} \citep{Antonucci_93,Urry_95}. Indirect evidence for the
existence of 
geometrically thick tori comes from the observation of broad emission
lines in the polarised flux of type~2 sources. First
detected in the nearby Seyfert~2 galaxy NGC~1068
\citep{Miller_83,Antonucci_85}, but later also in a sample
of other nearby sources \citep{Lumsden_04}, these observations were 
interpreted as arising from hidden type~1 nuclei.
Further indirect evidence and a first estimate of an average opening
angle of AGN tori comes from number statistics. 
\citet{Maiolino_95} find a ratio of Seyfert~2 to Seyfert~1 galaxies 
of 4:1 in their sample, which results in an opening angle of the
(more or less) dust-free cones of $74^{\circ}$. This is
in line with observations of light cones in nearby Seyfert galaxies.
Detailed SED fits for 513 type~1 AGNs from the XMM-COSMOS survey, 
allowed \citet{Lusso_13} to derive the ratio of re-processed mid-IR
emission to intrinsic nuclear luminosity, which is then converted into an
obscuration fraction. They find a shallow decrease of the obscuration
fraction with increasing AGN luminosity, which seems to be consistent
with the so-called {\it receding torus model} \citep{Lawrence_91,Simpson_05} and 
favours a model where the torus is optically thin in the mid-IR (see
also discussion in Sect.~\ref{sec:obscured}).

Due to their large distances and small sizes, AGN tori appear point-like in direct
infrared imaging
observations. Only with the advent of mid-infrared interferometry with
the help of the MIDI instrument \citep{Leinert_03}
at the Very Large Telescope Interferometer (VLTI) by pairwise combining the
light of the 8~m class VLT or 2~m class auxiliary telescopes they
could be resolved.
Very detailed observations have been carried out for the brightest
Seyfert galaxies NGC~1068
\citep{Jaffe_04,Poncelet_06,Raban_09,Lopez_Gonzaga_14} 
and the Circinus galaxy \citep{Tristram_07,Tristram_14}.
In these cases, the brightness distribution could be disentangled into
several components, which resemble an edge-on disc-like structure on sub-parsec
scales and an almost round, fluffy or filamentary larger component, probably tracing
the torus walls or dust in the torus funnel.
In a study of the brightest Seyfert~1 galaxy, NGC~4151,
\citet{Burtscher_09} find a similarly sized torus, whereas in the nearest
radio galaxy, Centaurus~A, most of the nuclear mid-IR emission seems
to be of non-thermal origin \citep{Meisenheimer_07,Burtscher_10}. 
Apart from NGC\,1068 and Circinus, elongations in polar direction (reminiscent of dust in the Narrow Line
Region, NLR) have been found in two additional objects, the Seyfert~2 galaxy
NGC~424 \citep{Hoenig_12} and the Seyfert~1.5 galaxy NGC~3783
\citep{Hoenig_13}.
The study of a larger sample of nearby AGN tori within the {\it MIDI AGN
snapshot survey} \citep{Tristram_09} and the {\it MIDI
AGN Large Programme} \citep{Burtscher_13}
revealed a diversity (but also a number of similarities) of complex, geometrically thick dust
morphologies on (sub-)parsec scale. 
With the help of near-infrared interferometry it is possible to
obtain the radius of hot dust close to its sublimation temperature,
which seems to scale with the square root of the AGN luminosity as
expected \citep{Barvainis_87,Suganuma_06,Kishimoto_11a,Weigelt_12}
and gives us a good measure of the inner radius of the torus.
For the case of sizes determined from mid-infrared (MIR) measurements,
the scaling with the AGN luminosity and its interpretation is still controversial
\citep{Tristram_09,Tristram_11,Kishimoto_11b,Burtscher_13}.

Due to the too low number of visibility measurements, direct image
reconstruction is currently not feasible for most sources.
Hence, for a careful data interpretation
and in order to get a more detailed idea of the structure of the torus,
dust continuum radiative transfer simulations for simplified
astrophysical scenarios or ad-hoc geometrical models have been carried out
and compared to high-resolution SEDs as well
as these interferometric observations. Mainly dictated by
computational requirements, they started off from simple geometries in
which the dust is distributed continuously
\citep{Pier_92b,Pier_93,Granato_94,Schartmann_05,Fritz_06}.
The latter were found to be in good agreement with the gross features of
observed SEDs, but continuous tori in
general showed too pronounced silicate emission features for type~1
sources.
Up to today, only weak silicate emission features have been found in various
sources showing AGN activity, ranging from very luminous quasars down
to weak LINERS \citep[e.~g.~][]{Siebenmorgen_05,Hao_05,Sturm_05,Weedman_05,Hao_07}.
This so-called {\it silicate feature problem} can be solved either by
changing the chemical composition of the dust
(e.~g.~\citealp{Fritz_06}) or by distributing the dust into
clouds. Both approaches can produce comparable strengths of the silicate
feature as has been shown by \citet{Feltre_12}. However, a
distribution of the dust in clouds has the advantage of being more
physical as it allows the dust to be 
protected from sputtering due to the surrounding hot gas
\citep[e.~g.~][]{Krolik_88}.
Being in reach with today's
  computational power, clumpy torus (toy) models have been set up
\citep{Nenkova_02,Hoenig_06,Nenkova_08a,Nenkova_08b,Schartmann_08,Hoenig_10a,Hoenig_10b,Stalevski_12},
which partly show good simultaneous agreement with spectroscopic as
well as interferometric observations.
The wealth and realism of the simulations mentioned above
impressively document our gain in understanding of geometrically thick 
AGN tori. However, several questions remain unanswered:
(i) How do tori build up and how do they evolve?
(ii) Which processes stabilise the scale height of AGN tori against gravity?
(iii) What is the geometrical structure and size of the torus?
(iv) Which physical processes generate the required clumpiness?

In early theories, the support was given by clump-clump collisions
with a high elasticity, mediated by strong magnetic fields
\citep{Krolik_88,Beckert_04}. 
It was also suggested to replace the torus by a magneto-centrifugally driven
disc wind solution, in which dusty clumps are embedded
\citep{Koenigl_94,Elitzur_06,Elitzur_09,Czerny_11} and form the
so-called {\it toroidal obscuration region} (TOR), which replaces the
classical torus.
Already \citet{Pier_92a} suggested that AGN tori can be supported by infrared
radiation pressure, after the ultraviolet/optical radiation from the
central source has been reprocessed within a thin layer at the inner
edge of the torus.  
This idea was successfully tested with the
help of analytical calculations in \citet{Krolik_07}, where they
simultaneously solve the radiative transfer problem as well as the
force balance. 
The effect of radiation pressure on single dusty torus clouds has been
studied in detail by \citet{Schartmann_11}, \citet{Plewa_13} and \citet{Namekata_14}.
The implications of a supernova-driven AGN torus is discussed in
\citet{Wada_02} and \citet{Wada_09}, where supernovae (SN) with a rate appropriate for
starburst conditions are invoked in a thin disc (following in-situ
star formation), which then 
puffs up the initially rotationally supported thin disc to form a
toroidal structure with obscuration properties as
observationally determined for AGN tori. 
The effects of the post-starburst evolution of a young nuclear star cluster
as ubiquitously found in nearby Seyfert galaxies \citep[e.~g.~][]{Davies_07} on the build-up
and evolution of AGN tori has been studied by 
\citet{Schartmann_09,Schartmann_10}, being able to explain the feeding
and obscuration of AGN tori. The processes at work typically form a
two-component structure: a filamentary large-scale torus and a
turbulent inner disc component, similar to the findings in the nearest
Seyfert galaxies revealed by MIDI. Subsequent radiative transfer
calculations enabled to connect this scenario to observations. 
Recently, another promising model for the build-up of AGN tori in
nearby Seyfert galaxies was presented by \citet{Wada_12}, which relies
on radiation feedback effects (see Sect.~\ref{sec:tor_model}). It is the aim of this article
to connect the density distribution resulting from this model to
observable quantities. We attempt to answer whether direct radiation
pressure and X-ray heating applied to a self-gravitating disc is able
to provide enough turbulent pressure to puff up and sustain AGN tori
and whether the structure is consistent with observed properties.
This is done with detailed radiative transfer calculations and
comparisons to high-resolution observations, where we are for the
first time able to provide time-resolved SEDs and images.
The parameters are chosen such to represent Seyfert activity
\citep{Seyfert_43}, as these are the best studied sources in the local
Universe. 

The hydrodynamical model itself will be recapitulated in more
detail in Sect.~\ref{sec:tor_model}. After describing the technical
methods and details in Sect.~\ref{sec:methods}, the results of the 
radiative transfer calculations are
characterised in Sect.~\ref{sec:results_rt}. 
We compare our findings in detail with available observations in
Sect.~\ref{sec:obscomparison}, 
followed up by a critical discussion (Sect.~\ref{sec:discussion}) and
conclusions are drawn in 
Sect.~\ref{sec:conclusions}.

\section{The underlying hydrodynamical model}
\label{sec:tor_model}

\begin{figure*}
\begin{center}
\includegraphics[width=0.75\linewidth]{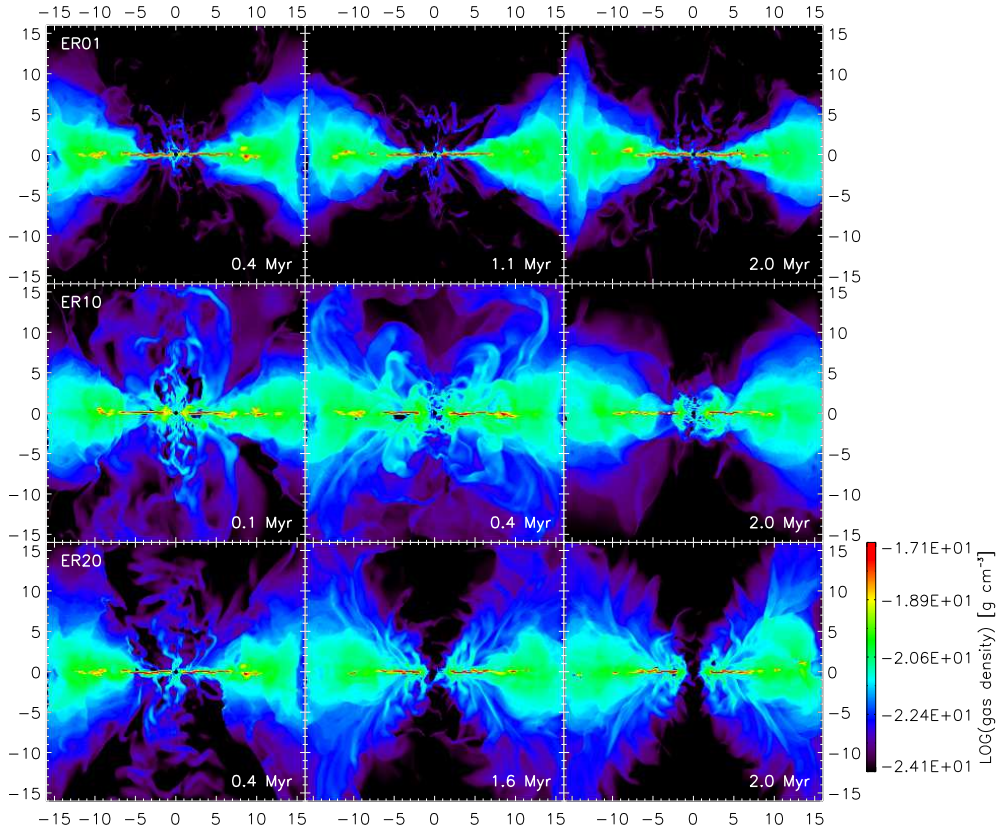} 
\caption{Cut through the y=0 plane of the gas density distribution for three characteristic time steps for
  models ER01 (upper panel), ER10 (middle panel) and ER20 (lower
  panel). A time of 0\,Myr corresponds to the switching on of the
  nuclear radiation. Labels are given in parsec. See
    \citet{Wada_12} for a detailed description of the dynamical evolution.}
\label{fig:rhogas_image}
\end{center}
\end{figure*}

\citet{Wada_12} proposed a physical model for the build-up 
of AGN tori with the help of X-ray heating and radiation pressure 
on dusty gas. To this end, \citet{Wada_12} ran three-dimensional hydrodynamical simulations 
coupled with the mentioned radiation processes, which are treated
with the help of a ray-tracing method
and include a $|cos(\theta)|$ radiation characteristic,
  mimicking the emission from a thin accretion disc. Additionally, they include 
the self-gravity of the gas, radiative cooling as well as
photoelectric heating due to uniform background UV radiation as well
as $H_2$ formation and destruction.
The mass of the central SMBH is chosen to be $M_{\mathrm{BH}} =1.3
\times 10^7\,\mathrm{M}_{\odot}$ (see
Tab.~\ref{tab:model_params}). Together with a time-independent
spherical external potential (mimicking the stellar distribution),
this results in rotation curves as observed in nearby Seyfert galaxies
(see \citealp{Wada_09} for a more detailed description). 
 
Starting with a marginally gravitationally unstable gas disc which has
already settled into a flared disc shape (with a
total gas mass of the evolved disc of $1.1 \times 10^6\,\mathrm{M}_{\odot}$), they
find that the radiation powers a vertical circulation of gas and dust
(``a radiation-feedback-driven fountain'') on parsec to tens of
parsec scale, which leads to a toroidal shape of the turbulently moving
obscuring material.  
In the simulations presented in this article, the radiation powers a
loss of gas from the computational domain with a rate of
$0.1\,\mathrm{M}_{\odot}\,\mathrm{yr}^{-1}$ on 
average. Hence, in order to keep the structure geometrically
puffed-up over typical AGN lifetimes of several tens of Myr, a mass
inflow rate from larger scales of the same magnitude 
is required, which is in the observable range of inflow rates within
bars and nuclear spirals \citep{Davies_14,Storchi_Bergmann_14}. 
Turbulent viscosity within the equatorial plane of the simulated discs
then replenishes gas in the innermost region, powering the circular flow
pattern.
The result of these processes is a multi-phase interstellar medium made
up of at least three components: (i) hot, diffuse ionised gas, in
which clumps and filaments of (ii) warm
molecular gas and (iii) dust are embedded.
A smaller Eddington ratio of the central source in
general leads to less
mass-loaded outflows and hence smaller amounts of gas falling back onto
the torus and hence causes less turbulent motions and a geometrically
thinner structure.
For the purpose of our radiative transfer calculations, 
resimulations of the two models in \citet{Wada_12} have been
carried out, namely for an
Eddington ratio of 1 per cent (model ER01) and 10 per
  cent (ER10).  A
third model with an Eddington ratio of 20 per cent
(ER20)\footnote{ We note that the Eddington ratios given
    here do not take the radiation characteristic of the central
    source into account. ``Effective'' Eddington ratios can be
    calculated by applying an angle average: 
   $\frac{\int^{\pi/2}_0  L_{\mathrm{AGN}} \cos(\theta) d\theta}{\int^{\pi/2}_0
     d\theta} = 2 / \pi$. 
    This then translates
    to Eddington ratios of 0.6 per cent (ER01), 6.4 per cent (ER10) and 12.7 per cent (ER20).} has been added.
The computational
domain
has been decreased by a factor of two, now simulating only a region
$32\,\mathrm{pc}^3$ around the centre, but keeping the number of grid cells identical
($256^3$). 
The simulation is calculated with self-gravity only until an
evolutionary time of 2.5\,Myr. 
Self-gravity in combination with thermal instability leads to
filamentary/spiral-like structures within the differentially rotating
disc \citep{Wada_01}. These structures
cannot survive over many rotational periods (0.1-1\,Myr), 
but similar structures are always reformed due to the instabilities since the
disc is kept at low temperatures by the radiative cooling. Therefore,
the entire filamentary morphology is long-lived over several Myr. 
After 2.5\,Myr, 
the radiation is switched on and
the first snapshot we use in this article is recorded after 0.1\,Myr. 
A total of 20 snapshots are used with a time
difference of 0.1\,Myr. 
Following \citet{Wada_12}, we translate the gas density to a dust
density distribution by applying a dust-to-gas mass ratio of 0.03 
\citep{Ohsuga_01} for all
gas with a temperature below $10^5$\,K. In regions with a gas
temperature above $10^5$\,K, the dust density is set to zero.
This is the temperature threshold where 
silicate and graphite grains get destroyed on short time scales,
caused by sputtering processes due to the surrounding hot gas
\citep{Dwek_96}.

\begin{figure}
\begin{center}
\includegraphics[width=0.85\linewidth]{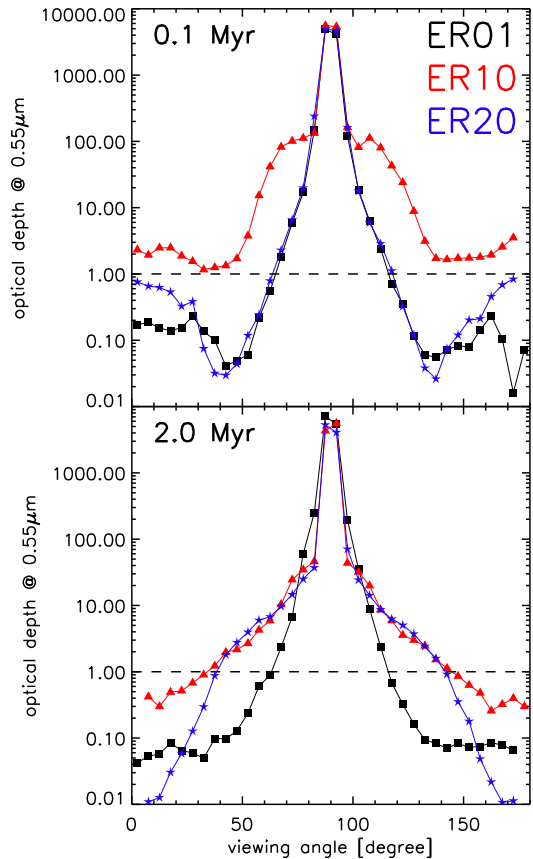} 
\caption{Optical depth at 0.55\,$\umu$m as a function of viewing angle,
  where $0^{\circ}$ and $180^{\circ}$ correspond to face-on lines of
  sight and $90^{\circ}$ to an edge-on view. Displayed are the models ER01 (black squares), ER10
  (red triangles) and ER20 (blue stars) at two time snapshots:
  0.1\,Myr (upper panel) and 2.0\,Myr (lower panel) after switching on
  the central radiation source. The dashed lines show the
  limiting optical depth used to determine the obscured fraction in
  Fig.~\ref{fig:obs_frac_direct_sed}.}
\label{fig:theta_tau}
\end{center}
\end{figure}

Fig.~\ref{fig:rhogas_image} shows a sequence of time snapshots for
meridional slices through the gas density distribution of
the three models (ER01 -- first row, ER10 -- second row and ER20 --
third row). There are already clear differences in the appearance
visible: model ER01 is not able to puff up the central few parsec
regime significantly; only an intermittent, low-density outflow
builds up. Model ER10 generates a geometrically thick structure on
parsec scale already early on (0.1\,Myr after switching on the radiation
source) including an outflow with a relatively high
mass-loading factor. Already after 0.4\,Myr, the outflow ceases
and the central region remains gas and dust enshrouded until the end
of the simulation.
The void conical structure at the latest
displayed snapshot is a consequence of the inflow motion at larger
distances from the midplane beyond the central gas enshrouded region.
In contrast, model ER20 is able to create a steady outflow with a very low gas and
dust density within the outflow-cone. However, the intermittent nature
of the outflow and energy feedback respectively is visible in the
filamentary, outward-moving structures.
The latter have velocities of the order of 100\,km\,s$^{-1}$ and
therefore remain within the computational domain for roughly 0.1\,Myr.
As soon as the model reaches a steady state, they are constantly
formed in the central region.
We cut out a central sphere with a radius of 1\,pc for the 
radiative transfer calculations, necessitated by the applied smoothing of the
gravitational forces in the central few cells in the hydrodynamical
calculation (see discussion in Sect.~\ref{sec:discussion}). 
The behaviour described above is quantified by displaying 
the optical depth at 0.55\,$\umu$m as a function of viewing angle (face-on
to edge-on). 
To this end, we determine the optical depth along 1000
randomly chosen radial rays. The average of these rays per viewing
angle bin of $5^{\circ}$ is shown in Fig.~\ref{fig:theta_tau} for the
first time snapshot (upper panel) and the last one (lower panel).
Whereas model ER01 shows the characteristic distribution for a thin
disc for all snapshots, both models ER10 and ER20 develop into a thick toroidal
structure for viewing angles close to edge-on viewing. Different from
model ER10, which develops a highly mass-loaded and later ceasing outflow from the
beginning, model ER20 is able to create a low-density outflow cone.  

\section{The radiative transfer modelling approach}
\label{sec:methods}

For our dust continuum radiative transfer calculations, we make use of
the versatile three-dimensional Monte-Carlo-Code 
{\sc RADMC-3D}\footnote{\url{http://www.ita.uni-heidelberg.de/~dullemond/software/radmc-3d/}}. 
We directly apply it to the three-dimensional cartesian grid described in
Sect.~\ref{sec:tor_model}. Apart from
the ability to calculate dust continuum radiative transfer, which is
used in our simulations, it also includes modules for gas line
transfer and gas continuum transfer.   
We use a typical galactic dust model as described in
\citet{Schartmann_05}, where the dust is split into 5 different grain sizes
(following an MRN size distribution, see \citealp{Mathis_77})
for the three different grain species each: silicate and the
two orientations of graphite grains with optical properties adapted
from \citet{Draine_84}, \citet{Laor_93} and \citet{Weingartner_01}.
The dust is heated by the central source, which is modelled as
a point-like emitter with a 
$|cos(\theta)|$ radiation characteristic, where $\theta$ is the polar 
angle counted from the rotation axis. The assumed spectral shape of the
central source is identical to the one used in \citet{Schartmann_05},
which was constructed from a combination of observations and
expected cut-offs from theoretical models for accretion discs.
After a thermal Monte-Carlo simulation \citep{Bjorkman_01,Lucy_99},
in which anisotropic scattering is treated using the Henyey-Greenstein
approximate formula, 
the resulting dust temperature distribution is used to calculate
images at various wavelengths as well as SEDs.
Given the large necessary inner radius (see Sect.~\ref{sec:tor_model}), not all
dust grains reach their sublimation temperatures. This especially has
some consequences for the near-IR (NIR) emission from the dust structure, as
discussed in detail in Sect.~\ref{sec:discussion}.
The basic parameter settings of our radiative transfer models as well
as the underlying hydrodynamical models are given in
Table~\ref{tab:model_params}. 

\begin{table}
\caption[Model parameters of the simulations.]{Model
  parameters of the simulations.}
 \label{tab:model_params}
\begin{tabular}{lrlr}
\hline
$M_{\mathrm{BH}}$ & $1.3 \times 10^7\,\mathrm{M}_{\odot}$ &
$M_{\mathrm{gas,ini}}$ & $1.1 \times 10^6\,\mathrm{M}_{\odot}$ \\
$\epsilon_{\mathrm{Edd}}$ & 0.01 (ER01) & $L_{\mathrm{source}}$ & $4.2\times 10^{9}\,\mathrm{L}_{\odot}$ \\ 
  & 0.1 (ER10) & & $4.2\times 10^{10}\,\mathrm{L}_{\odot}$ \\
  & 0.2 (ER20) & & $8.5\times 10^{10}\,\mathrm{L}_{\odot}$ \\ 
$f_{\mathrm{dtg}}$ & 0.03 & $T_{\mathrm{gas,thresh}}$ &  $10^5$\,K \\
$n_{\mathrm{lam}}$ & 300 & $n_{\mathrm{phot}}$ & $10^8$  \\
$n_{\mathrm{phot,scat}}$ & $10^6$ & $n_{\mathrm{phot,spec}}$ & $10^4$ \\
$x_{\mathrm{max}}$ & 16\,pc & $R_{\mathrm{in}}$ & 1\,pc \\
\hline
\end{tabular}

\medskip
$M_{\mathrm{BH}}$ is the central black hole mass, 
$M_{\mathrm{gas,ini}}$ is the total gas mass at the time the radiation
is switched on, 
$\epsilon_{\mathrm{Edd}}$ is the Eddington ratio, 
$L_{\mathrm{source}}$ the corresponding luminosity of the
central source, 
$f_{\mathrm{dtg}}$ the dust-to-gas ratio, $T_{\mathrm{gas,thresh}}$
the threshold gas temperature for dust to survive,
$n_{\mathrm{lam}}$ the number of wavelengths for the calculation of
the temperature distribution as well as the SEDs, $n_{\mathrm{phot}}$
the number of photon packages used in the thermal Monte-Carlo run, 
$n_{\mathrm{phot,scat}}$ the number of photon packages to calculate
the scattering in images (this number has been increased for the case
of the images at $0.1\,\umu$m to $10^8$), $n_{\mathrm{phot,spec}}$ the
number of photon packages for the calculation of scattering in the
SEDs, $x_{\mathrm{max}}$ the half side length of the computational
box and $R_{\mathrm{in}}$ is the inner radius for the radiative
transfer calculation. For a detailed description of the model
parameters of the hydrodynamical simulations, we refer to \citet{Wada_12}. 
\end{table}

\section{Results of the radiative transfer simulations}
\label{sec:results_rt}

\subsection{The wavelength dependent appearance of the obscuring structure}
\label{sec:tor_appearance}

\begin{figure*}
\begin{center}
\includegraphics[width=0.75\linewidth]{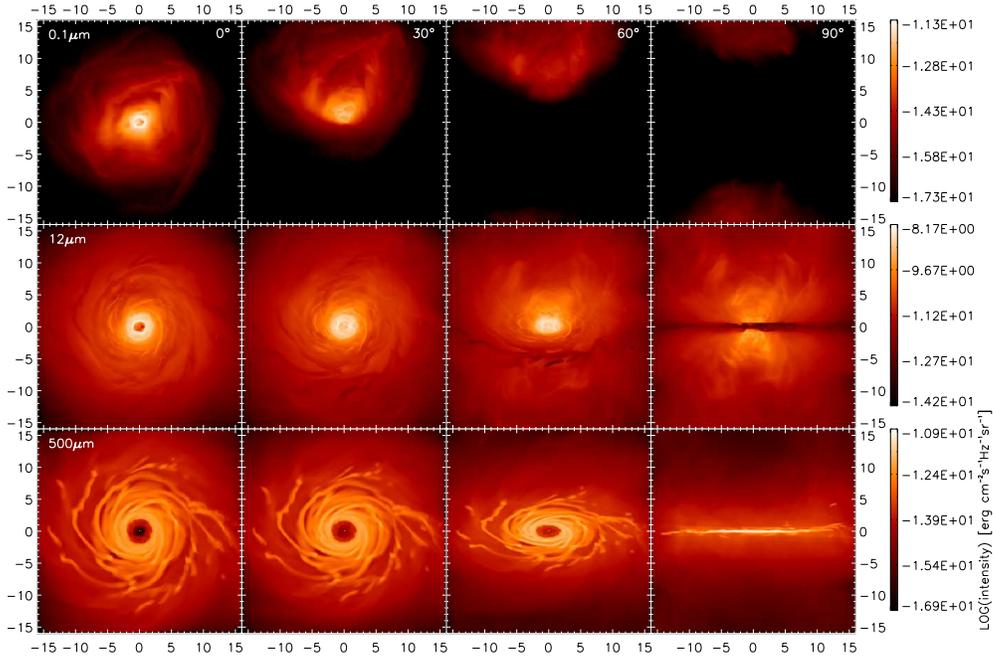} 
\caption{Wavelength dependence of model ER20 at a time 2.0\,Myr after
  the central radiation has been switched on. The various columns
  correspond to inclination angles of $0^{\circ}$, $30^{\circ}$, $60^{\circ}$ and
  $90^{\circ}$ (from left to right). The rows correspond to
  wavelengths of $0.1\,\umu$m (upper row), $12\,\umu$m (middle row) and 
  $500\,\umu$m (lower row). The dynamic range
    is chosen to scale from the maximum intensity of dust emission
    (excluding the central source) down to the $10^{-6}$th fraction of
    it. Labels are given in parsec.  
  }
\label{fig:wavelength_image}
\end{center}
\end{figure*}

\begin{figure*}
\begin{center}
\includegraphics[width=0.75\linewidth]{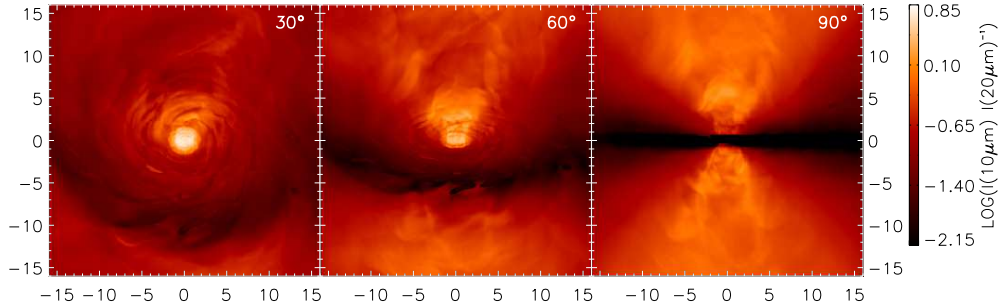} 
\caption{Ratio of intensity maps at a wavelength of $10\,\umu$m and $20\,\umu$m 
  -- indicative of dust temperature -- of model ER20 at a time 2.0\,Myr after
  the central radiation has been switched on, where brighter colours
  correspond to higher temperatures. The panels
  correspond to inclination angles of $30^{\circ}$, $60^{\circ}$ and
  $90^{\circ}$ (from left to right). The dynamic range
  is chosen to scale from the maximum ratio
  down to the $10^{-3}$rd fraction of
  it. Labels are given in parsec.}  
\label{fig:image_ratio_10_20}
\end{center}
\end{figure*}

Fig.~\ref{fig:wavelength_image} shows the inclination angle 
(from face-on -- column 1 to edge-on -- column 4 in steps of
$30^{\circ}$) and wavelength ($0.1\,\umu$m -- upper row, $12\,\umu$m --
middle row and $500\,\umu$m -- lower row) dependence of the ER20 model
at a time of 2.0\,Myr.
Looking first at the mid-infrared images (middle row) one can clearly
see that it changes from a centrally symmetric structure to something more and
more asymmetric. At higher inclinations,
high-column-density filaments from close to the equatorial plane
become visible in form of dark absorption bands and partly end up on
the line of sight. Already at $i=30^{\circ}$ they can be seen against the
background of the emission from the low density
dust in the funnel region and its directly illuminated walls. The latter becomes the
dominant structure in the edge-on and 
near edge-on view, where it turns into an X-shaped feature.
The appearance is significantly different at short wavelengths. Shown in the
upper row of Fig.~\ref{fig:wavelength_image} is the intensity at
$0.1\,\umu$m, corresponding to the maximum emission of the central
source. The visible 
structures are dominated by scattered light as well as absorption, as
this is also the peak of the opacity model used (see
Sect.~\ref{sec:methods} and compare to Fig.~3a in
\citealp{Schartmann_05}). Due to the high absorption  
efficiency, only scattered light from the low-density cones is visible.
Please mind the
orders of magnitude difference in the intensity compared to the other
wavelengths shown (see colour bars).
The lowermost row
displays images at $500\,\umu$m. In this case, the emission is
dominated by dust at very low
temperatures. Additionally, the opacity has dropped dramatically in
this wavelength regime. 
Hence, the largest fraction of the obscuring structure
itself is optically thin and only the high-density disc with
spiral-like features in the
midplane of the simulation shows up. 
Such structures are ubiquitous in imaging observations and 
dust extinction maps (Sect.~\ref{sec:imobs}).
The emerging multi-component structure can best be seen in 
the ratios of the brightness distributions at $10\,\umu$m and
$20\,\umu$m, which is displayed for three inclination angles in 
Fig.~\ref{fig:image_ratio_10_20}.
The hot dust (brighter at $10\umu$m) is concentrated close to the
centre, as well as within the low-density outflow cones. 
Towards the midplane, the disc gets colder and colder and dominates
the longer wavelength emission (darker colours).
This multi-component nature of the central gas and dust distribution
is clearly not very well captured by the term ``torus''.  Therefore,
we will replace it in the following by ``obscuring structure'' or
similar.

\subsection{Time evolution of mid-infrared images}
\label{sec:mir_images}

\begin{figure*}
\begin{center}
\includegraphics[width=0.75\linewidth]{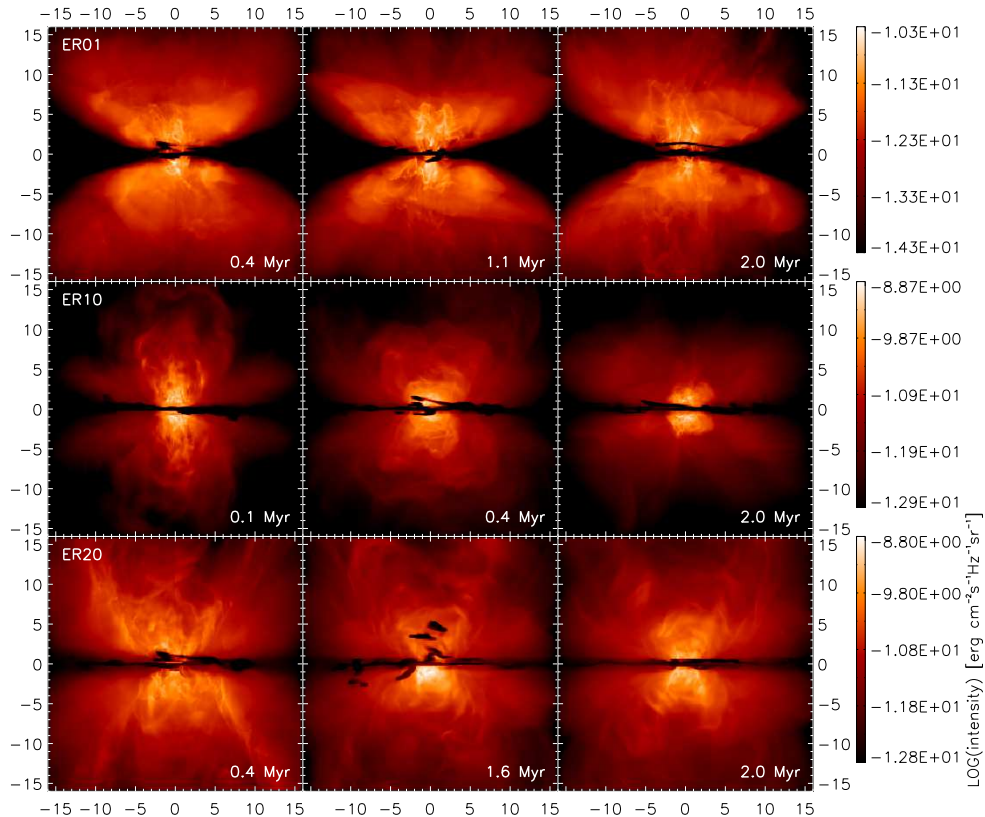} 
\caption{MIR images at $12\,\umu$m of the three characteristic time
  steps shown in Fig.~\ref{fig:rhogas_image} for
  models ER01 (upper panel), ER10 (middle panel) and ER20 (lower
  panel). Shown is the edge-on view ($i=90^{\circ}$). Labels are given
in parsec.}
\label{fig:evolution_image}
\end{center}
\end{figure*}

The same timesteps as shown in the gas density distributions in
Fig.~\ref{fig:rhogas_image} and discussed in
Sect.~\ref{sec:tor_model} are shown in 
Fig.~\ref{fig:evolution_image}. 
As already detailed above, for the case of model ER01 (upper row), the radiation
is not able to significantly increase the disc scale height. Only a
filamentary, low-density outflow arises, which clearly shows up in the
$12\,\umu$m images. Due to the low densities above and below the disc
and the non-existence of a small scale obscuring structure,
the emission cone has a comparatively large opening angle.
The dense filamentary disc shows up in form of a central absorption
band, strengthened by the
applied $|\cos\theta|$ radiation characteristic.
Model ER10 (middle row) represents an example, which starts with a
high column density 
filamentary outflow (see Fig.~\ref{fig:theta_tau}), leading to a vertically elongated appearance of
the edge-on view onto the dust distribution. At later snapshots, the outflow ceases, resulting in a
temporary X-shaped emission (middle panel) and later in a centrally concentrated and
spherically symmetric morphology {(right panel).
The model ER20 (lower row) shows the strongest outflow, visible in the
formation of a low density, but inhomogeneous funnel and hence the funnel
walls can be more or less directly heated from the central source,
showing up as X-shaped features (see Sect.~\ref{sec:imobs}), especially in the first displayed
snapshot. At later times, the initially thin disc gets puffed-up more and more
leading to a decreasing opening angle of the funnel.
In some snapshots, the
absorption due to high density clumps leads to an asymmetric appearance
of the dust configuration. An example can be seen in the middle panel.

\subsection{Time-resolved spectral energy distributions}
\label{sec:seds}

\begin{figure}
\begin{center}
\includegraphics[width=0.95\linewidth]{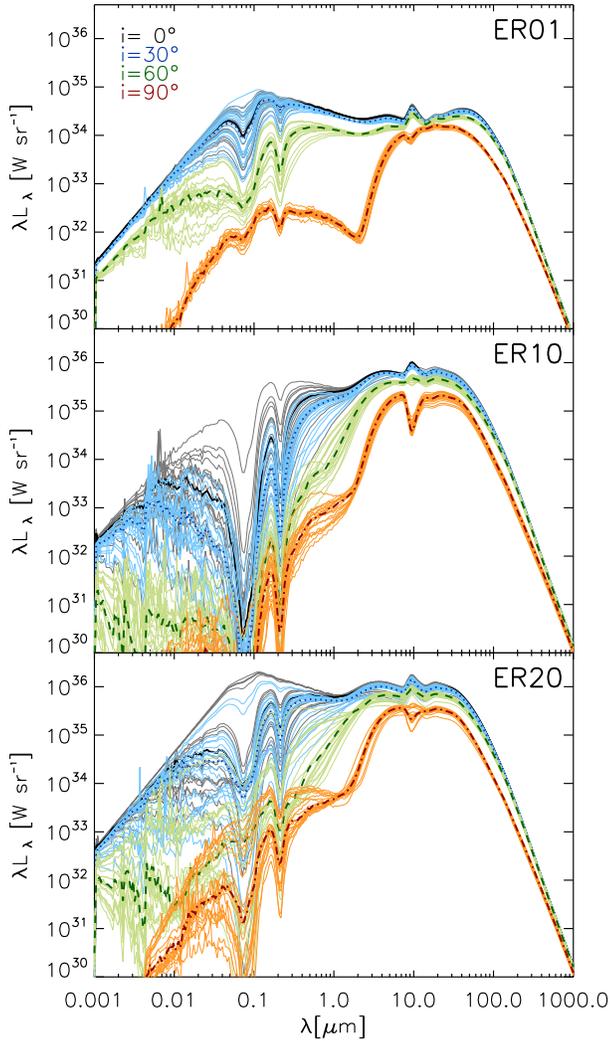} 
\caption{Time evolution of the SEDs of model ER01 (upper panel), ER10
  (middle panel) and ER20 (lower panel). 
  The thick lines represent the median of all time
  snapshots of the respective models at inclinations of  
  $0^{\circ}$ (black solid), $30^{\circ}$ (blue dotted), $60^{\circ}$
  (green dashed) and $90^{\circ}$ (red dash-dotted). SEDs of the
  individual snapshots are given in lighter colour.}
\label{fig:sed_comparison}
\end{center}
\end{figure}

Fig.~\ref{fig:sed_comparison} shows a compilation of SEDs 
of all models and time steps discussed in this article. 
The different panels
refer to the three Eddington ratios taken into account: from 1 per cent
(upper panel, ER01) to 10 per cent
(middle panel, ER10) and 20 per cent (lower panel, ER20).
Inclinations are colour-coded (black solid - $i=0^{\circ}$, face-on,
blue dotted -
$i=30^{\circ}$, green dashed - $60^{\circ}$, red dash-dotted - $90^{\circ}$, edge-on),
where the thin lines in light colour refer to the
individual SEDs from the time series and thick lines show the SEDs
averaged over the available time steps.
In general two bumps are visible: (i) the ``Big Blue Bump'' -- peaking at 
roughly $0.1\,\umu$m -- and (ii) the ``IR bump'', peaking at around $10\,\umu$m. 
The first represents the SED of the central source, which acts 
as the only source of energy in our simulations (see
Sect.\,\ref{sec:methods} and Fig.~3b in \citealp{Schartmann_05} for
details). It is partially absorbed due to dust on the line of sight and
slightly altered by scattering off dust grains.
Heating the surrounding 
dust to temperatures close to the sublimation temperature results
in re-emission of the grains in the infrared wavelength
regime, visible in the form of the ``IR bump''.
Overlaid over the two bumps are spectral features:  most prominent
and also most important for our analysis is the {\it 10 micron feature} 
(peaking more accurately at $9.7\umu$m, see discussion in
Sect.~\ref{sec:shi_plot_comparison}),
arising due to stretching modes in
silicate tetrahedra (Si-O stretching modes).
Only slightly visible is the second resonance of silicate dust,
namely the 18.5\,$\umu$m feature (attributed to O-Si-O bending modes).
The very pronounced absorption feature at around $0.08\,\umu$m has 
contributions both from the small graphite grains and small silicate
grains. At $0.2175\,\umu$m, graphite shows a significant feature.
Given the temperature constraints at these wavelengths, the latter only
show up as absorption features against the background of the flux from
the central source SED. As the dust extinction
properties as well as the central source SED
peak at roughly $0.1\,\umu$m, dust gets heated very efficiently to temperatures
up to the sublimation temperature (approximately
1500\,K for graphite grains and 1000\,K for silicate grains). 
When inclining the toroidal structure from face-on view towards
edge-on view (black solid $\to$ blue dotted $\to$ green dashed $\to$
red dash-dotted), more and 
more cold dust moves into the line of sight towards the nucleus. The
resulting absorption is most visible around the peak of the opacity
curve (~$0.1\,\umu$m, see e.g. \citealp{Schartmann_05}) and the
graphite (and silicate) resonances show up as absorption feature. 
Also the overall emission in
the IR bump decreases due to the increasing absorption along the
line of sight. The biggest difference is visible between 60$^{\circ}$ and
90$^{\circ}$ inclination angle, due to the remaining large column
densities within the equatorial plane when looking exactly edge-on
onto the dust distribution.
This effect is strongest for model ER01, as expected for a geometrically
thin disc. ER10 shows the deepest absorption features, caused by the 
compact density distribution of the failed wind. Model ER20 results in
the largest differences at short wavelengths, caused by the already
mentioned three-component structure: disc, obscuring structure and outflow.
It is also noteworthy to mention that model ER20 shows a variety of
feature strengths at the various time steps for the edge-on case
(see the zoomed-in view of the SEDs in Fig.~\ref{fig:sed_zoomin_er20}): 
The silicate feature changes from slight emission (partly self-absorbed) during the early evolution
(black to blue thick lines) when the structure is still disc-dominated, to strong
absorption during the high-density outflow phase (cyan intermediately
thick lines) to the final
steady state with moderate absorption (green to red thin lines),
which is reached after roughly 1.6\,Myr (compare to the
  light curve shown in the lower panel of Fig.~\ref{fig:lightcurves_paper}).
The variation of the silicate absorption feature in the quasi-steady
state gives a hint towards the clump size distribution or the
contribution of hot dust within the funnel. 
The majority of our modelled SEDs show emission features, even for
inclinations as high as $60^{\circ}$ (see discussion
in Sect.~\ref{sec:shi_plot_comparison}).

\begin{figure}
\begin{center}
\includegraphics[width=0.95\linewidth]{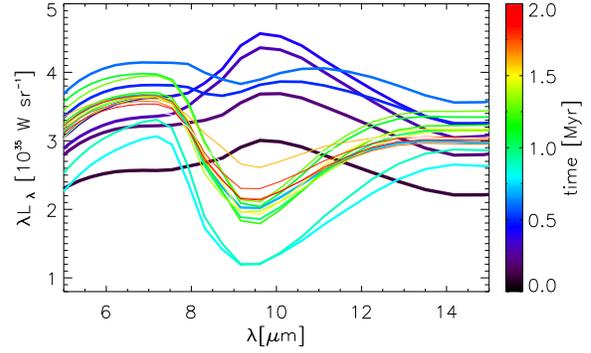} 
\caption{Time evolution of the silicate $9.7\,\umu$m feature in the SEDs of model
  ER20 for an inclination angle of $90^{\circ}$. 
  The line colours refer to various evolutionary stages
    as indicated in the colour bar. Line thickness decreases from the
    first snapshot to the last.}
\label{fig:sed_zoomin_er20}
\end{center}
\end{figure}

\subsection{Light curves}
\label{sec:lightcurves}

\begin{figure}
\begin{center}
\includegraphics[width=0.95\linewidth]{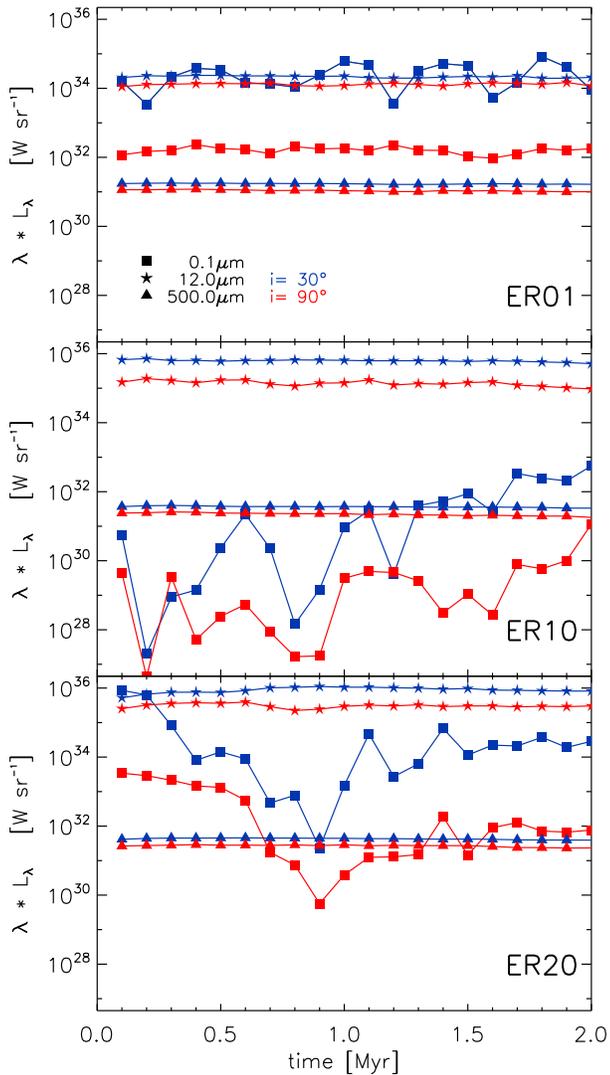} 
\caption{Light curves for the models ER01 (upper panel), ER10 (middle
  panel) and ER20 (lower panel) for wavelengths of $0.1\,\umu$m (squares), $12.0\,\umu$m
(stars) and $500.0\,\umu$m (triangles) at inclination angles of
$30^{\circ}$ (blue) and $90^{\circ}$ (red).}
\label{fig:lightcurves_paper}
\end{center}
\end{figure}

Fig.~\ref{fig:lightcurves_paper} shows light curves for the three
models for wavelengths of $0.1\,\umu$m (squares), $12.0\,\umu$m (stars) and
$500.0\,\umu$m (triangles) at inclination angles of
$30^{\circ}$ (blue) and $90^{\circ}$ (red).
The $500.0\,\umu$m (triangles) light curve traces the cold dust
component, e.~g.~the dense disc only. As the latter remains in a
quasi-steady state, a constant
time evolution can be observed for all three models during the
2\,Myr time frame. At this wavelength, the obscuring structure is optically thin for
all calculated inclinations
  ($0^{\circ},30^{\circ},60^{\circ},90^{\circ}$), except for 
a slightly lower luminosity for the edge-on view.
This is the expected result for a thin disc.
For the case of the $12.0\,\umu$m light curve (stars), model ER01 shows
the same behaviour, but for the other two models, the flux
  decrease with inclination is slightly stronger, as expected. 
At the shortest probed wavelength of $0.1\,\umu$m (squares), the
differences between the inclinations are the highest, because this is at
the maximum of the opacity curve. At these wavelengths only scattering
off dust grains can act as a source term apart from the primary
radiation. The relatively constant time
evolution of model ER01 is indicative of the low density (and hence
optical depth) of the lifted dust. Furthermore, no strong clumpiness
exists in this model. 
In model ER10, the outflow ceases relatively early on, 
but compared to
the other two models, no steady state solution is reached. Instead,
during the further evolution,
the column density towards the poles
slightly decreases with time (see Fig.~\ref{fig:theta_tau}), which is
reflected in the rising trend of the short wavelength light curve.
Finally, this behaviour might lead to a periodical
  re-establishment of an outflow along the axis, as can be seen in the
time-evolution of the same model in \citet{Wada_12}.
In contrast, model ER20 shows a
dip, correlated with an episode of strong and dense outflows, 
emphasizing the intermittent behaviour of the model.

\subsection{Evolution in the colour-colour diagram}
\label{sec:evol_colcoldia}

\begin{figure}
\begin{center}
\includegraphics[width=0.95\linewidth]{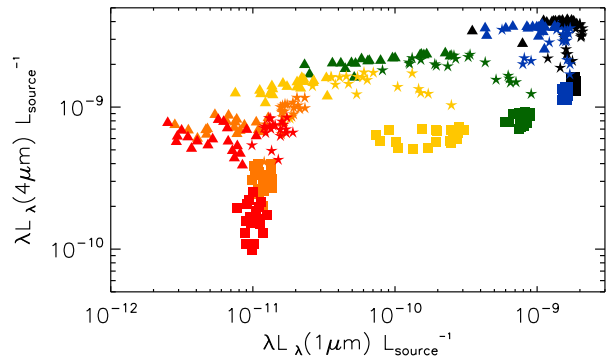} 
\caption{Colour-colour diagram for models ER01 (squares), ER10
  (triangles) and ER20 (stars) at inclination angles of
$0^{\circ}$ (black), $30^{\circ}$ (blue), $60^{\circ}$ (green) and
$90^{\circ}$ (red). Intermediate inclination angles are given in
yellow ($70^{\circ}$) and orange ($80^{\circ}$). 
All 20 time steps are
shown for each model and inclination.}
\label{fig:colcoldia}
\end{center}
\end{figure}

Fig.~\ref{fig:colcoldia} shows a colour-colour diagram, where the
emission at $4\,\umu$m is compared to emission at $1\,\umu$m. 
Model ER01 (represented by the squares) clearly shows distinct
concentrations for every inclination angle (due to the
  applied radiation characteristic), with only
  small time variations: strong $1$ and $4\,\umu$m emission for inclinations up to 
$60^{\circ}$ and at low luminosities for both wavelengths for
near edge-on views.
Between $60^{\circ}$ and $90^{\circ}$, first the $1\,\umu$m flux
decreases (yellow squares, $i=70^{\circ}$) and later also the $4\,\umu$m flux
(orange squares, $i=80^{\circ}$), dictated by the shape of the
opacity curve. In total it resembles a broken
power law distribution. This is the expected 
signature for a geometrically thin disc with a sharp,
discontinous evolution from a type~1 to a type~2 source. 
For the case of models ER10 and ER20, a similar behaviour is seen, but
shifted towards higher $4\,\umu$m fluxes. In a geometrically thick 
structure, a larger fraction of the primary radiation intercepts with
dust and is re-emitted in the IR. The maximum flux at $1\,\umu$m is
roughly the same for all models, due to the normalisation by the total
input luminosity. The distribution of the individual inclinations of
the two models along the x-axis is much more stretched compared to
ER01, which reflects the stronger time-dependence of those models 
resulting in a real fountain process with higher density clumps
embedded. It is, however, strongest for ER10 which is caused by the
intermittency of the outflow, finally resulting in a failed wind, which
enshrouds the nucleus.

\section{Comparison with observations}
\label{sec:obscomparison}

After having characterised the observable appearance of the models, 
we will do a first comparison with available observational data in the
following.

\subsection{Imaging observations of nearby Seyfert galaxies}
\label{sec:imobs}

Recent imaging observations of the best
studied nearby low luminosity galaxies with ground-based 8m-class and
space-based telescopes reveal 
striking similarities to 
our simulated images.
These studies reach the currently highest possible resolutions and
typically image the tens of parsec surrounding of the central SMBH.
To this end, they are not directly comparable in size to our
simulations, but a continuation of the structures found there 
might be expected, as we will motivate in the following.
The mentioned X-shaped morphology at MIR wavelengths, which is 
mainly seen for model ER20 in Fig.~\ref{fig:evolution_image} (lower row) can be 
recovered in the observations of the Circinus galaxy analysed by
\citet{Prieto_04}, see especially their multi-wavelength image (their Fig.~2).
Here, the V-shaped structure marks the transition to the
{\it one-sided ionisation cone}. The observed
central K-band source would then correspond to our central, directly 
illuminated core region.
A similar {\it tongue}-like appearance is also found in the parsec-scale
vicinity of NGC~1068 in form of the so-called UV cloud B
\citep{Evans_91}, compare also to Fig.~3 (bottom left) in
\citet[][hereafter P14]{Prieto_14}.
In these cases, the low extinction within the strong,
but low-density outflow (as found in model ER20) enables the scattered
light to be visible (Fig.~\ref{fig:wavelength_image}, upper row) and the escape
of coronal lines. A multitude of such {\it ionisation cones} have been
observed in nearby sources.
They are supposed to be caused by the main obscurer on
(sub-)parsec-scale and hence expected to extend from (sub-)parsec up
to hundreds of parsec scale. Indeed, dust emission along the direction
of the ionisation cones is also
detected on parsec-scales in the interferometric observations with the
MIDI instrument \citep{Tristram_14,Lopez_Gonzaga_14}.
The behaviour of our models (especially ER20) at wavelengths of
$0.1\,\umu$m and $12\,\umu$m at
inclination angles $i\ge 60^{\circ}$ (Fig.~\ref{fig:wavelength_image})
is reminiscent of the  
observed flux distribution of the Seyfert~2 galaxy MCG~05-23-016. 
Two cones of emission are visible in the optical
wavelength regime in combination with an extinction lane (see Fig.~2
mid left panel in P14). The latter is even better
visible in the dust extinction map as shown in their Fig.~2 (mid right
panel). 
The dust extinction maps are a convenient way of imaging the
morphology of warm dust with respect to the background galaxy light.
In this sense, they can -- from a morphological point of view -- be 
compared to our emission maps of the warm dust, e.~g.~at 12$\,\umu$m.   
Spiral-like features and filaments -- as revealed in our long wavelength images
(Fig.~\ref{fig:wavelength_image}, lower row) -- are ubiquitous
structures in the cores of nearby active galaxies,
which have recently also been detected in dust extinction maps, but on
slightly larger scales (dictated by resolution constraints).
Prominent examples are NGC~1097 (see Fig.~5 in \citealp{Prieto_05})
and ESO~428-G14 (see Fig.~1 in P14).
In summary, similar filamentary structures to the ones predicted by
our simulations in the parsec-scale vicinity of the central
engine are already observed in extinction maps on the smallest
resolvable scales (corresponding to the tens of parsec central regions
of nearby Seyfert galaxies). Future observations will be able to test
our predictions for even smaller scales.

\subsection{The average nuclear SED in Seyfert galaxies}
\label{sec:av_sed}

\begin{figure}
\begin{center}
\includegraphics[width=0.95\linewidth]{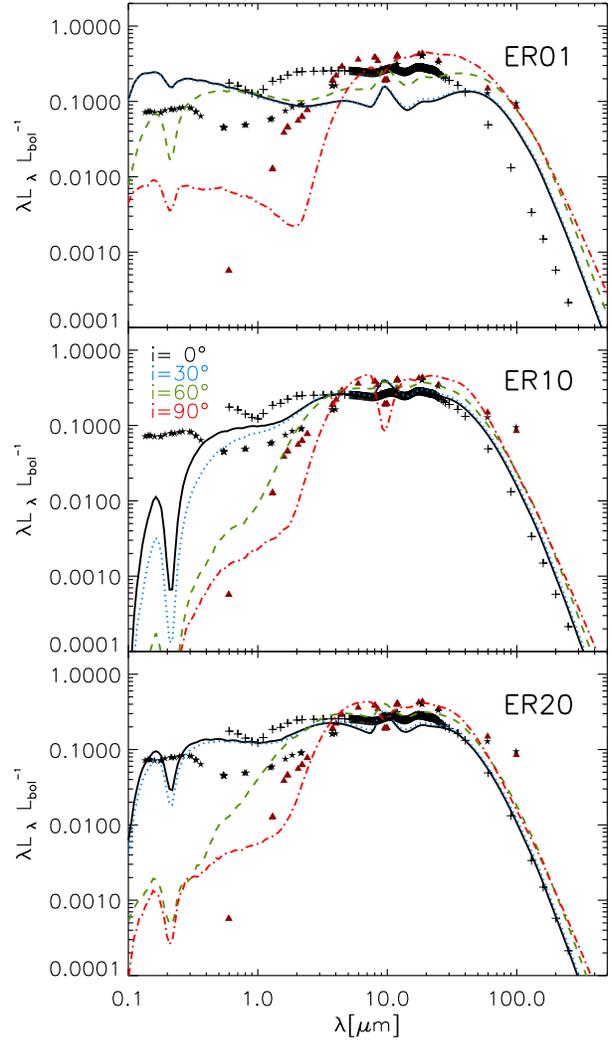} 
\caption{Comparison of model SEDs to the observed templates for Seyfert~1 galaxies
(black stars) and Seyfert~2 galaxies (red triangles) of
\citet{Prieto_10} as well as the composite type~1 SED of
\citet{Mor_12}, given by the black plus symbols. 
All SEDs are normalised to their total bolometric luminosity
$L_{\mathrm{bol}}$, but are otherwise the same 
as the ones shown in Fig.~\ref{fig:sed_comparison}: 
ER01 (upper panel), ER10
  (middle panel) and ER20 (lower panel) at inclinations of  
  $0^{\circ}$ (black solid), $30^{\circ}$ (blue dotted), $60^{\circ}$
  (green dashed) and
  $90^{\circ}$ (red dash-dotted). Only the median SEDs of all time
  snapshots of the respective models are shown.}
\label{fig:sed_comparison_template}
\end{center}
\end{figure}

Fig.~\ref{fig:sed_comparison_template} shows the time-averaged SEDs of all models
normalised to their total bolometric luminosity $L_{\mathrm{bol}}$
(see Fig.~\ref{fig:sed_comparison} for the original SEDs). 
This is done in order to enable a comparison with
high spatial resolution observed templates for Seyfert~1 (black stars)
and Seyfert~2 (red triangles) galaxies \citep[][hereafter P10]{Prieto_10}.
The latter have been normalised to their total emission of the IR and 
UV bumps as well, within the wavelength regime between
0.1 and $100\,\umu$m.
Although the Seyfert templates rely on low-number statistics (three
type~1 sources and four type~2 sources are included in the average,
see P10) and a large variety of
spectral shapes are observed, they give us the best current estimate
of the core SED of Seyfert galaxies -- i.~e.~medium to high luminosity AGNs. 
Due to their
high spatial resolution, they represent the intrinsic AGN plus central dust
SED and, hence, allow a direct comparison with our modelling.  
We additionally show the composite SED from \citet[][hereafter MN12]{Mor_12}, which has
been recently extended towards shorter wavelengths (Netzer, private
communication) as the black plus symbols. It is in good agreement with 
the \citet{Elvis_94} template, with only larger deviations at
far-infrared (FIR) wavelengths.
The MN12 template has also been normalised to its total luminosity within its
available wavelength range (0.6-250\,$\umu$m).
It is an average of 51 narrow-line Seyfert\,1 galaxies (NLS1s) and 64
broad-line Seyfert\,1 galaxies (BLS1s) with a large variety of
luminosities, observed with {\it Spitzer}.
Given the large beam size, the star formation contribution has been
subtracted with the help of star formation templates, taking into
account the full range of possible host galaxy properties.
This template is slightly bluer compared to the one of P10 (black
stars), with the largest deviations corresponding to roughly a factor of two higher
luminosities in the NIR.
We attribute the latter
to a combination of a higher level of extinction along the
line-of-sight of the
Seyfert~1 galaxies in the P10 sample and to the
contribution of luminous objects (PG quasars) that may dominate the
MN12 sample in this wavelength regime.
As expected, model ER01 neither gives a good match to the type~1 nor the type~2
templates. 
Concerning the type~2 template, the models either
show too little extinction ($0^{\circ}$ to $60^{\circ}$) or a
deficiency of flux in the NIR at $90^{\circ}$. A good match
can only be expected for a very narrow range in inclination angle.
The models overpredict the
long wavelength emission and the silicate feature in absorption is not as
pronounced as in the observations.
A detailed comparison for the type~1 sources is more difficult due to
the discrepancy of the observed composite SEDs. However, an extinction 
comparable to the sources in the P10 template can only be achieved for
inclinations larger than $60^{\circ}$, which is
unrealistic for type~1 nuclei. In addition, the
models are neither in good agreement with the MN12 template, as they show
a deficiency of emission in the NIR.
The models also tend to produce too much emission at longer
wavelengths. The reason for this behaviour is the morphology of the model, which is
dominated by the thin and dense disc (see Sect.~\ref{sec:tor_model}).

A better match with the data is found for model ER10. 
However, most of the time snapshots show more extinction than
observed at short wavelengths ($\lambda<0.3\umu$m), even for the 
case of the P10 type~1 template. 
The fully edge-on view 
shows on average a too pronounced silicate absorption feature. 
This behaviour is expected for this model, as -- due to the 
failed outflow -- the central core is fully enshrouded with dust,
leading to the highest column densities in polar direction of all
models (Fig.~\ref{fig:theta_tau}). 

Model ER20 results in a reasonably good comparison of the edge-on
models with the observed type~2 template. The apparent discrepancy at
wavelengths around $2\,\umu$m (and related to this the uncertainty in
this wavelength regime due to our choice of a large inner radius) is
mitigated by the fact that the scatter 
of the SEDs combined in the type~2 template is larger than the visible
{\it knee} structure, which is bracketed by the simulated SEDs for $60^{\circ}$ and $90^{\circ}$
inclination. Surprisingly, even the
silicate absorption feature depth is roughly reproduced on average. 
At the largest simulated wavelengths (where the obscuring structure is
more or less optically thin), the SEDs show too little flux compared to the P10
templates, but are in rough agreement with the MN12 template.
The differences in the templates in this wavelength regime most probably arise from
differences in the treatment of the subtraction of starburst
contributions, spatial resolution as well as the sample. 
As already mentioned, the P10 average SED is made up of Seyfert galaxies
only,
the MN12 one is dominated by quasars, which on average
have higher Eddington ratios compared to local
Seyfert galaxies. This also appears to be reflected in our Eddington ratio
study, giving a better comparison of the ER20 model to the MN12
template and resulting in a closer match of the low Eddington ratio
models to the P10 template.
Concerning Seyfert~1 observations,
the majority of the ER20 models seem to reproduce the correct absorption 
at around 
$0.2\,\umu$m and show NIR emission in-between the 
P10 and MN12 templates.
However, a clear deficiency of our models is the emergence of deep
graphite features at $0.2\,\umu$m, which are never seen in observations. This is an
unsolved problem related to the dust model used in our simulations.
The NIR is also the spectral range where our models suffer from
the largest uncertainties. First of all, our relatively large inner
radius does not allow all dust grain species to reach their
sublimation temperature on the one hand. On the other,
the cut out dust does
not contribute to extinction and clumpiness on small scales,
which causes a larger fraction of hotter dust in the outflow cone. 
The latter also depends on the assumed 
dust-to-gas fraction, which might be spatially varying and  
significantly reduced in the hot outflow cones.
Additionally, in this region, the choice of the central source
spectrum can also play a role. 
All three model families show too strong
silicate emission features (see also discussion in 
Sect.~\ref{sec:shi_plot_comparison}). 

We conclude that the overall best adaptation of the observations is given by model ER20, leading 
to the conclusion that a three-component structure made up of a high
column density thin disc plus a geometrically thick surrounding
atmosphere, which collimates a low density outflow, 
is capable to explain these observational constraints and replaces the
classical ``torus'' in this model.

\subsection{The silicate feature strength / column density relation}
\label{sec:shi_plot_comparison}

\begin{figure}
\begin{center}
\includegraphics[width=0.95\linewidth]{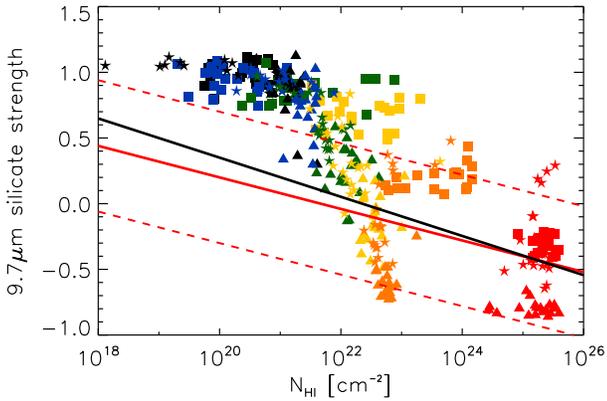} 
\caption{
  The silicate feature strength plotted as a function of the neutral
  hydrogen column density
  for model ER01 (squares), ER10 (triangles) and ER20
  (stars) at inclination angles of $0^{\circ}$ (black), $30^{\circ}$
  (blue), $60^{\circ}$ (green), $70^{\circ}$ (yellow), $80^{\circ}$ (orange) and $90^{\circ}$ (red). 
  The fit to an observed sample of active galaxies is
  shown as the black line, whereas the red line is a fit to the
  sub-sample of Seyfert galaxies \citep{Shi_06}.
  The dashed red lines are meant to guide the eye concerning the
  observed scatter around the relation for the Seyfert galaxies.}
\label{fig:feat_colden}
\end{center}
\end{figure}

In Fig.~\ref{fig:feat_colden}, the silicate feature strength
is shown as a function of the neutral hydrogen column density. 
To calculate $N_\mathrm{{\textsc HI}}$, only cells with
a gas temperature below the ionisation temperature of hydrogen ($1.7\times10^4\,$K) are
integrated along the given line of sight.
The strength of the silicate feature is determined in analogy to
\citet{Shi_06} as
\begin{eqnarray}
\Delta_{\mathrm{feature}} = \frac{F_{\mathrm{feat}}-F_{\mathrm{cont}}}{F_{\mathrm{cont}}},
\end{eqnarray}
where $F_{\mathrm{feat}}$ is the flux at the wavelength of the
silicate feature (taken where the maximum deviation from the continuum
is reached) and $F_{\mathrm{cont}}$ is the spline fitted
continuum at the same position, anchored at wavelengths between $5.0$
and $7.5\,\umu$m and from $25.0$ to $40.0\,\umu$m. 
These wavelength ranges are chosen such to be outside
the two broad silicate features at $9.7\umu$m and $18.5\,\umu$m. 
Being partly in the decreasing region of the SED, this choice slightly
affects the determined silicate feature strengthes, but we consider it
the cleanest option.
The distribution of observed AGNs in this diagram has been determined
by \citet{Shi_06}. They compiled a sample of 85 AGNs of various types
for which Spitzer IRS \citep{Houck_04} observations or ground-based
measurements from the \citet{Roche_91} sample were available as well
as {\sc HI} column density information from X-ray spectra through the {\it
  Chandra} data archive or from literature. 
The resulting relationship for the full sample is shown by the
black line, whereas the red line depicts the observed relation for the
Seyfert galaxy sub-sample. Both show significant scatter (roughly
accounted for by the red dashed lines, but compare to Fig.~3 in
\citealp{Shi_06} for the actual observed distribution) which was
interpreted as clumpiness of the absorbing gas and dust distribution. 
Overplotted in Fig.~\ref{fig:feat_colden} is the resulting
distribution for our models. The inclination of the simulations is colour-coded
($0^{\circ}$ -- black, $30^{\circ}$ -- blue, $60^{\circ}$ -- green,
$70^{\circ}$ -- yellow, $80^{\circ}$ -- orange, $90^{\circ}$ -- red)
and the various symbols refer to the three models (ER01 -- squares, ER10
-- triangles and ER20 -- stars).
For model ER10, we find a too steep slope with only a small amount of
scatter up to an inclination angle of $80^{\circ}$, followed by a jump
in column density. The reason for this is again the failed wind,
leading to a geometrically thick shell of gas surrounding the central
region. Such a configuration fails to explain the observed trend,
as was also found in the analysis of smooth torus models
\citep{Schartmann_05}, which was done in Fig.~11 of
\citet{Schartmann_09}. 
The jump in column density (but not in silicate feature
strength) can be explained by the thin but very dense disc component
in all the models. Whereas the neutral column density probes only the
line of sight, the silicate feature strength is a superposition of the
effects from the full model space (depending on optical depth along
the individual lines of sight).
For the same reason, the edge-on views for models ER01 and ER20 --
although having similar column densities -- end up on or
close to the observed
relation in contrast to ER10. Whereas the scatter is too small for
ER01, it is comparable to the observed scatter for the higher
Eddington ratio model ER20. The reason for this can be found in the
stronger time variation / intermittency of ER20 (which in the steady state is
identical to a rotation of the obscuring structure along the symmetry axis), caused
by the more filamentary / clumpy nature of this model (see
Fig.~\ref{fig:rhogas_image} and \ref{fig:evolution_image}).

All of the models produce too strong silicate emission features
compared to the observed relation. 
The best adaptation of the data is given by model ER20, which seems to
follow a broken powerlaw distribution, providing a good representation
of the observations at large inclination angles, but a too steep slope
at small inclination angles. The latter is related to the fact that
the warm gas emission seems to be too extended above the high density
region of the obscuring structure and probably missing clumpiness in the central
region (see discussion in \citealp{Schartmann_08}), which might be
related to too low resolution (0.125\,pc) of the underlying hydrodynamical
simulations in this central part. Another reason for the discrepancy 
might also be related to the observations. Given the low spatial
resolution of the Spitzer observations, the true nuclear emission feature strength
of the central dusty component heated by the central source can also
easily be diluted due to additional
flux contributions from within the beam, e.~g.~caused by the host
galaxy or a circumnuclear star burst. Concerning the absorption feature 
strength, the problem is confusion caused by additional dust extinction
due to dusty filaments or clumps along the line of sight, as is e.~g.~the case in Centaurus~A, where a
prominent circumnuclear dust lane might cause additional extinction
(see also discussion in Sect.~\ref{sec:imobs}, P14 and \citealp{Goulding_12}).

\subsection{The obscured fraction}
\label{sec:obscured}

\begin{figure}
\begin{center}
\includegraphics[width=0.95\linewidth]{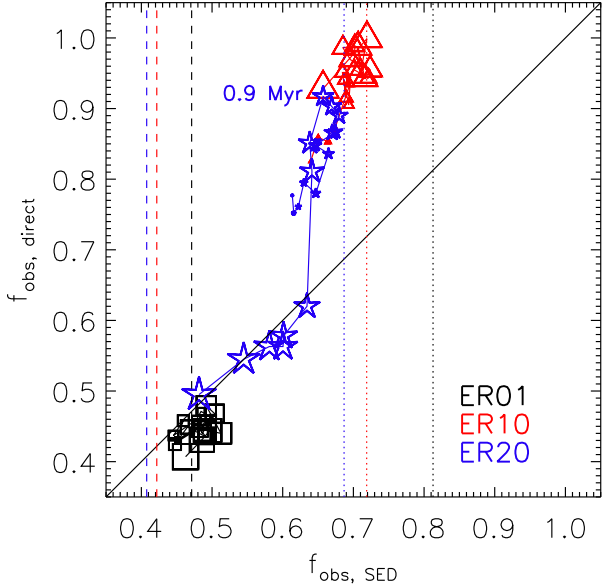} 
\caption{Obscured fraction for the models ER01 (black squares), ER10
 (red triangles) and ER20 (blue stars) as determined from the SEDs plotted
  against the directly determined obscured fraction.
  The symbol size shrinks with time between the first (0.1\,Myr)
  and the last recorded snapshot at 2.0\,Myr.
   The vertical lines are the values for the
   three (effective) luminosities determined
   from a fit to observations by \citet{Lusso_13} for the assumption of
   an optically thin (dotted lines) and optically thick obscuring structure
   (dashed lines) calculated with Eq.\,\ref{eq:obsfrac}.
}
\label{fig:obs_frac_direct_sed}
\end{center}
\end{figure}

In this subsection, we introduce two methods for
  determining the obscured fraction of active galactic nuclei, which
  we subsequently apply to our set of simulations. Firstly, we
use 1000 randomly chosen lines of sight to
directly determine the
optical depth at $0.55\,\umu$m along radial rays (see Fig.~\ref{fig:theta_tau}). 
The obscured fraction is then given by the ratio of the number of
rays with an optical depth larger than one (as indicated by the
horizontal line in Fig. ~\ref{fig:theta_tau}) to the total number of
rays and is shown in Fig.~\ref{fig:obs_frac_direct_sed} 
along the y-axis. 
The second method we use follows the approach of \citet{Lusso_13} (as
already mentioned in Sect.~\ref{sec:introduction}) and derives the
obscured fraction from the SEDs. The results of this derivation are
given along the x-axis of Fig.~\ref{fig:obs_frac_direct_sed} and will
be discussed below. Concerning the first method, all data from 
model ER01 (black squares) -- resembling the thin disc initial condition for the whole
simulation time -- show the expected very low obscuration fraction. 
For the case of model ER10 (red triangles), the obscuration fraction
scatters around a high, constant value and later decreases
(symbols get smaller with increasing time). This
matches well the increasing trend of the $0.1\,\umu$m light curve
(Fig.~\ref{fig:lightcurves_paper}, middle panel,
  squares) . The correspondence of the obscured fraction to the light curve 
is even more evident for model ER20 (blue stars) with 
a maximum in $f_{\mathrm{obs,direct}}$ at the same time ($t=0.9\,$Myr)
as the minimum in the short wavelength ($0.1\,\umu$m) light curve (see
Fig.~\ref{fig:lightcurves_paper}, lower panel, squares). 

The obscured fraction as a function of nuclear luminosity has been
observationally examined by \citet{Lusso_13}. By fitting the observed broadband 
SED and decomposing it into starburst, hot dust re-emission, host
galaxy and big blue bump components,
they are able to determine the ratio $R$ of hot dust re-emission
luminosity and the
intrinsic AGN luminosity ($R=L_{\mathrm{IR}} / L_{\mathrm{UV}}$),
which directly relates to the obscured fraction of AGN by \citep{Granato_94}

\begin{eqnarray}
  f_{\mathrm{obs}} \simeq \frac{R}{1+R(1-p)},
\end{eqnarray} 

where p is an anisotropy parameter defined as the ratio between the
integrated emission of the dust in edge-on view and face-on view. 
Using 513 type~1 AGNs from the XMM-COSMOS survey, \citet{Lusso_13} are able to
fit the resulting luminosity dependence of the obscured fraction 
by a receding torus model with luminosity dependent scale
height ($h\propto \xi$, see \citealp{Simpson_05}):

\begin{eqnarray}
  \label{eq:obsfrac}
  f_{\mathrm{obs}} = \left[ 1+3\,\left( \frac{L_{\mathrm{bol}}}{\mathcal{L}_0} \right)^{1-2\,\xi}\right]^{-0.5}
\end{eqnarray} 

where $\mathcal{L}_0=10^{46.16}\mathrm{erg}\,\mathrm{s}^{-1}$ and
$\xi=0.37$ for the assumption of the obscuring structure to be
optically thin in the mid-infrared ($p = 1$) and $\mathcal{L}_0=10^{42.65}\mathrm{erg}\,\mathrm{s}^{-1}$ and $\xi=0.44$ for
optically thick obscuring structures in the mid-infrared ($p \ll 1$).
As our central source is anisotropically emitting, we
  replace $L_{\mathrm{bol}}$ with the angle averaged quantity:

\begin{eqnarray}
  L_{\mathrm{bol}} = \epsilon_{\mathrm{Edd}} \, \frac{2}{\pi} \, L_{\mathrm{source}}
\end{eqnarray} 

The resulting obscured fractions from this formula are shown in Fig.~\ref{fig:obs_frac_direct_sed} as the
vertical dotted (dashed) lines in the respective colours for the assumption of
optically thin (thick) obscuring structures, clearly demonstrating the very shallow
correlation with luminosity. The directly determined obscuring
fractions (Fig.~\ref{fig:obs_frac_direct_sed}, plotted
  along the y-axis) are in reasonable agreement with the observationally
determined values for the cases of ER10 and ER20 in the later stages
of the evolution. Both models finally
approach the observations for the optically thin case
($f_{\mathrm{obs}}=0.74$ for ER10 and $f_{\mathrm{obs}}=0.71$ for ER20), corresponding to the
interpretation favoured by \citet{Lusso_13}. However, one has to
take into account that the luminosity change in our modelling approach
is solely caused by a change of the Eddington ratio and not the black
hole mass itself. Additionally we are investigating the low luminosity end of the
observed relation, which is not very well sampled observationally. Hence, model ER01
cannot directly be compared to the observed relation (extrapolated fraction),
but it shows very good correspondence to the observed value for the
optically thick interpretation ($f_{\mathrm{obs}}=0.48$ for ER01).
We should also note that as we fix the inner radius to 1\,pc,
independent of the sublimation distance in all
models, the behaviour cannot be related to the receding torus
model. 

Fig.~\ref{fig:obs_frac_direct_sed} also shows the obscured fraction
when determined in the same way as in \citet{Lusso_13}, but taking the
anisotropy parameter into account. According to
this analysis, all models are in agreement with the observations, as
they lie within the limiting cases of optically thick
($f_{\mathrm{obs}}=0.48$ for ER01, $f_{\mathrm{obs}}=0.43$ for ER10 and
$f_{\mathrm{obs}}=0.42$ for ER20) and thin ($f_{\mathrm{obs}}=0.83$ for ER01, $f_{\mathrm{obs}}=0.74$ for ER10 and
$f_{\mathrm{obs}}=0.71$ for ER20) obscuring structures in AGNs. 
Whereas model ER01 is consistent with the  
observations assuming an optically thick obscuring structure for mid-IR emission,
model ER10 coincides with the observations, interpreted as an
optically thin obscuring structure.
Whereas model ER01 gives roughly 
consistent results for both methods, this is only the case in the beginning of the
evolution for model ER20. The directly determined fraction is higher
compared to the one determined from the SEDs for model ER10 and in the 
equilibrium state of model ER20. 
This clearly shows the difficulties of the
interpretation of the obscuring fraction with the help of SEDs
for the case of our models.

\section{Discussion}
\label{sec:discussion}

Recent detailed investigations of a sample of nearby Seyfert galaxies
observed with MIDI reveals a diversity of obscuring structures in AGNs
\citep{Burtscher_13}. Whereas a fraction of the sources show clear
signs of elongations (mostly in polar direction for the best-studied
cases), others can be well explained by a symmetric brightness
distribution. A similar behaviour is not only found in
the Eddington ratio study presented in this work
(Fig.~\ref{fig:evolution_image}), but already in the
dynamical evolution of the single models with time. One can easily imagine that an
additional variation of the (remarkably few) model parameters (e.~g.~the black hole
mass, see Table~\ref{tab:model_params}) will add to this diversity in appearance.
Hence, a strict one-to-one comparison with observations is not useful
at this point, but we mainly restrict ourselves to comparing to
average observed properties.

There is a striking similarity of the filamentary structure of the
thin disc component of the model
(Fig.~\ref{fig:wavelength_image}) with the filaments and spiral-like
patterns found in the extinction maps for nearby Seyfert galaxies
\citep[][P14]{Prieto_05}, but on larger scales.   
These are the typical features caused by disc instabilities
\citep[e.~g.~][]{Toomre_64,Wada_01,Behrendt_14}, which enable
the gas to be fed towards the central parsec scale region.

One caveat of the radiative transfer simulations used in this paper is the comparatively
low resolution (0.125\,pc) of the underlying hydrodynamical simulations in the central region. This does not allow to resolve
the small-scale structure there, which leads to a smearing out of the
density distribution and might be partly responsible for the deviations seen
in the comparison to the observed silicate feature strength versus
column density relation. We were therefore forced to set a large
inner radius of one parsec, which is outside the expected dust
sublimation radii for most of the grain species we use and hence no grain
dependent sublimation can be applied in these simulations. 
After a detailed analysis of the temperature distributions, we find
that in model ER01 none of the dust reaches above our assumed
sublimation temperatures (1000\,K for silicate and 1500\,K for
graphite dust). In model ER10, a fraction of $8.3\times 10^{-7}$ of
the total dust mass is up to a factor of 1.3 above the sublimation temperature and in model
ER20 this amounts to a fraction of $2.9\times 10^{-6}$ with a maximum temperature
up to a factor of  1.5 above the sublimation temperature.
The grain dependent sublimation was found to influence especially 
the near-IR emission properties (most important for the low
Eddington ratio model ER01, in which the nominal dust sublimation
radius is well below the inner radius) as well as the strength of the
silicate emission feature of the models \citep{Schartmann_05} and
should be further elaborated in upcoming high-resolution simulations.  
To this end, a quantitative and detailed comparison to interferometric measurements
of the two nearest Seyfert galaxies NGC~1068 and the
Circinus galaxy is currently not feasible, 
but this should be done with the next generation physical models,
due to the much better constraining power compared to a sole comparison
to SEDs. This is even more the case for the upcoming Matisse instrument
\citep{Lopez_12} given its imaging and multi-band capabilities.

\section{Conclusions}
\label{sec:conclusions}

Recently, \citet{Wada_12} proposed a physical model for the evolution
of the gas and dust distribution within AGN cores,
where X-ray heating in combination with radiation pressure
on dust leads to the establishment of a {\it radiation-driven fountain}
process in an initially thin disc. A vortex motion of the gas is created, where the back-flowing
part drives turbulent motions in the disc, which significantly
increases its scale height. The idea of this article is to connect
this physical model with observational properties. To this end we feed
the resulting time-dependent gas distribution into the Monte-Carlo dust radiative
transfer code {\sc RADMC-3D} in order to calculate images and spectral
energy distributions (SEDs), which we compare to available
observations. This enables us for the first time to derive the time
evolution of the dust emission from an obscuring structure around an AGN. The parameters are
chosen such to resemble nearby Seyfert galaxies, which are the nearest
and hence best observed active galaxies.
The input density distributions are characterised by a three-component
structure: (i) a very dense, but geometrically thin disc, which is
structured by gravitational instability into clumps, which get
stretched into filaments by tidal forces and (ii) a surrounding
atmosphere -- responsible for most of the obscuration -- with significant sub-structure formed by
column density instability mediated by radiation pressure, which
collimates a (iii) low density outflow along the rotation axis. 
This multi-component morphology is also reflected in the temperature structure,
strengthened by the applied $|\cos(\theta)|$ radiation characteristic. 
The latter leads to dramatic differences when observing at different
wavelengths: the appearance in most cases changes from a vertically
elongated structure -- tracing the outflow cone -- at short
wavelengths to a disc-like appearance at longer wavelengths. 
The intermittent behaviour of the flow is directly visible in the
image series in the mid-IR as well as indirectly in the SEDs as well,
which naturally show the strongest time-dependence at short
wavelengths close to $0.1\,\umu$m where both the source spectrum as
well as the opacity curve has its maximum. 
From a study of various Eddington ratios for the central luminosity we
favour those which create a strong and fast outflow, with low column
density close to the poles and a wide opening angle of the obscuring
material (model ER20 for our sample). 
These typically show mid-IR images which are elongated in
polar direction (partly displaying X-shaped features), 
which seems to be in agreement with recent findings
with the help of the MIDI instrument for a number of nearby Seyfert galaxies.  
However, when comparing to the silicate feature strength versus column density
relation, we find good agreement at high column densities, but too
strong silicate emission features at low inclination angles and column
densities. The latter being related to too much emission from dust within
the funnel or from the funnel walls (partly with unfavourable morphology) or
by a too small amount of clumpiness, which might be caused by too low
resolution in the nuclear region close to the radiation source or the
missing grain dependent sublimation due to a too large inner radius.  

In summary we find good agreement of the proposed physical model for
the build-up and evolution of obscuring structures in AGNs
with available observations.
In future, we plan to compare the revealed three-component structure (as found in the
best-fit model ER20) of a higher resolution simulation
also covering the sublimation region to high spatial resolution
interferometric observations in various wavebands.

\section*{Acknowledgments}

We are greatful to Hagai Netzer for helpful discussions and for
providing us with an extension of the composite SED presented in
\citet{Mor_12} towards shorter wavelengths prior to publication.
A detailed and constructive report by an anonymous referee
helped to improve the publication.
We thank C.~P.~Dullemond for making 
RADMC-3D\footnote{\url{http://www.ita.uni-heidelberg.de/~dullemond/software/radmc-3d/}}
publicly available and for his continuous support with the code.
This work was supported by the Deutsche Forschungsgemeinschaft priority program 1573
(``Physics of the Interstellar Medium''). KW is partly supported
by JSPS KAKENHI Grant Number 23540267.  AP thanks for the hospitality
of the Max Planck Institute for extraterrestrial Physics and the CAST group.

\bibliographystyle{mn2e}
\bibliography{abb_journals,literature}

\end{document}